\documentclass[11pt,fleqn]{article}
\usepackage{graphicx}
\begin{document}
\small

\noindent
\begin{center}
\textbf{\huge The mean density of the Universe from cluster evolution}\vspace{3ex}\\
Alain Blanchard, James G. Bartlett and Rachida Sadat\\
\vspace{1ex}
\textit{ Observatoire Astronomique, 
            11, rue de l'Universit\'e, 67000 Strasbourg, France 
          }
\end{center}

\noindent
\textbf{{\large Abstract}}\\
\vspace{1ex}

\noindent
{\footnotesize 
The determination of the mean density of the Universe is a long standing 
problem of modern cosmology. 
The number density evolution of x-ray 
clusters at a  fixed  temperature
is a powerful cosmological test,  new in nature (Oukbir and Blanchard, 1992),
 somewhat different from standard 
analyses based on the dynamical measurement of individual objects. However, 
the absence of any available sample of x-ray selected clusters with measured 
temperatures at high redshift has prevented this test from being applied earlier. Recently,  
temperature measurements of ten EMSS clusters at $0.3 \le z \le 0.4$ have 
allowed the application of this test (Henry, 1997). 
In this work, we present the first results of  
a new analysis we have performed of this data 
set as well as a new estimation of the local temperature distribution 
function of clusters:  a likelihood analysis of the temperature distribution functions gives a 
preferred value for the mean density of the universe which corresponds to 75\% of 
the critical density. An open model with a density smaller than 30\% of 
the critical density is rejected with a  level of significance of 95\%.}\\

\noindent
\section {Introduction} 
 
The value of the mean density of the Universe, ${\rho}_{0}$, is a fundamental 
constant for 
Cosmology. According to the theory of general relativity, in the case of a 
zero cosmological constant, the evolution
of the Universe is determined by the ratio of this density to the so-called 
critical density, ${\rho}_{c}$,  the latter corresponding to the solution known as the 
Einstein-de Sitter model.  Models 
with a density greater than this critical density will undergo a finite 
period of expansion, followed by a collapse sometimes called 
the ``big-crunch''.
On the contrary, if the density is smaller than the critical value,  the expansion will continue forever.
This is all conveniently parameterized by $\Omega_{0}$, the density 
{\em parameter}, which is equal to the ratio of the actual 
density to the critical density.  The value of $\Omega_{0}$ also has
implications for the local geometry of space, for general relativity
provides a link between the two. If  $\Omega_{0}$ is $>1$, the 
geometry is spherical; if $\Omega_{0}$ is $<1$, the geometry is 
hyperbolic; while in the special case of $\Omega_{0} = 1$, the geometry of 
space is locally Euclidean.  Finally, the value of $\Omega_{0} $
should be predicted by the theories of the very early Universe.
Inflation is certainly the best example of such a theory, whose initial 
prediction was $\Omega_{0} = 1$. It has been recently shown that other 
values are possible, but at the expense of a lack of simplicity.\\

The classic method of determining the density of the Universe is known as the 
Oort method. It consists of evaluating the amount of mass present in 
a specific 
object, like a cluster for instance, from knowledge of the velocity 
dispersion, or the x--ray gas temperature. The 
mean value of the universal density is then derived under the assumption that 
the ratio  of total to visible mass is constant. Such a method currently 
leads to low values of $\Omega$, of the order of $0.2-0.4$ (Adami et al., 1998). The reliability 
of this method has been a matter of long debate over the last fifteen 
years, and remains so to this day.
 
\section{Cluster number evolution: a global probe of 
 the mean density of the Universe} 

It is clear that it is vital to find methods to estimate the mean density of 
the universe which do not rely on this assumption of a constant ratio between 
mass and visible material. Geometrical tests, like the luminosity--distance 
relation, could in principle provide us with such a method. 
In practice, these tests have always failed to give a reliable answer
because they are pervaded by evolutionary effects. The application to distant 
supernova represents a modern version which has received much 
attention in recent years, possibly indicating a non-zero 
cosmological constant (Riess, et al., 1998).  Recently, 
it has been realized that the statistical properties of
the cosmological microwave bakcground fluctuations maps could 
allow one 
to determine the cosmological parameters with high precision (provided that 
the theoretical framework of gravitational instability with initial adiabatic 
fluctuations is correct). Although current observations already 
lead to interesting conclusions, only satellite--based
observations, like those from Planck or MAP (Microwave Anisotropy Probe), 
will allow to remove the degeneracy between the various parameters entering the problem.\\
A category of test which is different in nature, is based on the dynamics of perturbations 
in the expanding universe. Linear perturbations grow at a slower
on the dynamics of perturbations in the expanding universe.  Linear 
perturbations grow at a slower rate in a low density 
Universe than in $\Omega_0=1$ Universe.  It follows that 
knowledge of the cosmic velocity field would lead to the value of $\Omega_0$ (Dekel, 1994). First 
applications of this test were promising, but it has since been realized that 
systematic effects could present a stronger limitation to its
usefulness than was initially thought (Davis, 1998).
Another approach, proposed a couple of years 
ago, is based on the evolution of 
the number density of x-ray clusters (Oukbir and Blanchard, 1992). This evolution is directly 
related to
the growth rate of fluctuations and allows, in principle, one to measure the 
mean density of the Universe.  Considerable work has been devoted
to the study of this test (Bartlett, 1997, and references therein). 
These studies show unambiguously 
that the abundance
of x-ray clusters, with temperatures of the order of few keV, is expected to 
change much  more 
slowly with redshift in a low--density Universe than in an Einstein--de Sitter 
model, a strong difference that should be easy to measure. 
\\

The results from two x-ray satellites, ASCA and ROSAT, 
have considerably improved the quality 
of the data: the luminosity function of x-ray clusters 
now seems well established (Ebeling et al., 1997), 
and the number of x-ray clusters at high 
redshift with measured temperatures has substantially increased. This has 
allowed a first indirect application of this test to a sample of high--$z$ x-ray clusters,
leading to a rather high density  for the Universe, close to the critical 
value (Sadat et al., 1998). Finally, 
Henry (1997) has for the first time provided the temperature of 
a set of clusters at significantly high redshifts ($ 0.3 < z < 0.4$), 
selected in a well-defined way, thereby allowing an estimation of
the temperature distribution function. 
In the following, we present the estimation of the mean density of the Universe
on the basis of this set of clusters and of a new set of local clusters 
($ z \sim 0.05$).  

\section{Data Analysis} 

We use a compilation of ROSAT observations of clusters 
and select those above 
a flux limit of 
$2.2 10^{-11}$ erg/s/cm$^2$, leading to a sample of fifty clusters. 
Temperatures for these fifty clusters 
were taken from the literature. Although it is believed 
that this sample is complete, it remains possible that some clusters 
have been missed. In such a case, the value of the mean density of the 
universe we infer would 
be {\em underestimated}. The temperature distribution function can be 
estimated following the standard method:
$$
N(>T) = \sum 1/V_m
$$
where $V_m$ is the volume of the sample out to the maximum depth at 
which the cluster would have been detected, given its intrinsic luminosity 
and the flux limit of the sample. 

An important source of possible bias in the estimation of the 
temperature distribution
function comes from the temperature measurement errors 
(Eke et al., 1998; Viana,  Liddle, 1998)~: because there are 
considerably more low temperature than high temperature clusters, this  
can produce an apparent cluster abundance which is higher than the 
actual value. This effect was early pointed by Evrard (1989)
for the velocity dispersion distribution function. 
In the present case, the 
errors differ significantly from one measurement to another and are 
correlated with the apparent luminosity, which determines the 
volume $V_m$. A correction to individual clusters is therefore preferable 
to a simple mean correction over the whole sample. We have therefore 
applied a Bayesian correction to individual temperature measurements. 
The main difference with 
previous estimation lies in the fact that we obtain a number density for 
clusters with $T \sim 4$ keV significantly higher. We have inferred 
the temperature distribution function 
at high redshift in a similar way by using 
the data as provided by Henry (1997).

\section{Estimation of the mean density} 

The method we chose to estimate $\Omega_0$ is the maximum 
likelihood estimate on the number density of x-ray clusters  
in {\em independent} bins of the temperature 
distribution function at $z = 0.05$ and $z = 0.33$. Such 
an analysis requires knowledge of the distribution function of the 
estimator of the mean number density of clusters. This distribution $p$
was found by using the bootstrap re-sampling technique. 
The likelihood function is then 
computed as :
$$
{\cal L} = \prod_{ i } p(N_i|N(\sigma_c,n,z_i,\Omega_0))
$$
where $\sigma_c$ is the amplitude of the fluctuations on cluster scales, $n$
is the power spectrum index of the primordial fluctuations, $z_i$ is 
the redshift of the i--th bin considered, $N_i$ is the actual observed 
number density of clusters
in this bin, while $N$ is 
number of clusters {\em predicted} by the model using the Press and Schechter 
mass 
function (Press and Schechter, 1974). The best estimate parameters correspond 
to those for which 
$\cal L $ is 
maximum. A 68\%, 95\%, confidence intervals on one parameter can be 
obtained by considering the 
region enclosed 
by $\Delta {\cal L} = 0.5 $ and $\Delta {\cal L} = 2. $, respectively
(this is only indicative, as it only holds for a normal distribution, 
which is not valid in our case). The maximum likelihood value we obtain is 
$\Omega_{0}=0.74$. The 95\% range, according to normal statistics, is 
[0.3 - 1.2] (symmetrized). 
A preliminary analysis of the various sources of systematic uncertainties
indicates that this number is not likely to change by a large amount.

\section{Conclusion} 

The method we have applied to determine the mean density of the Universe
has the considerable advantage of being global,  relying on 
the dynamics of the Universe as a whole. Furthermore, numerous studies 
have confirmed the power 
and robustness of this method.  The conclusion of a 
high density universe we have obtained, consistently with Sadat et al. (1998),
 could represent a 
major and fundamental advance in the understanding of our Universe; and, 
consequently, it calls for considerable prudence. The temperature 
distribution function we obtain at $z = 0.05$ is based on ROSAT fluxes  
which are believed to be accurate as well 
as on recent temperature measurements for the largest set available 
(fifty clusters). Great caution should be taken with 
the sample of high redshift clusters:   any un--identified 
systematic effect 
in the selection function could undermine our estimate. In the near 
future, two new spectro-imaging, x-ray satellites, AXAF and XMM,
will very likely bring much more light on the nature of distant clusters, 
allowing a definitive answer to these questions. It is therefore 
tempting to believe that the mean density of the Universe will be robustly 
determined before the end of the century.

\begin{figure}
~\vspace*{-1.5cm}
\centerline{\hspace*{1.cm}\resizebox{\hsize}{!}{\includegraphics{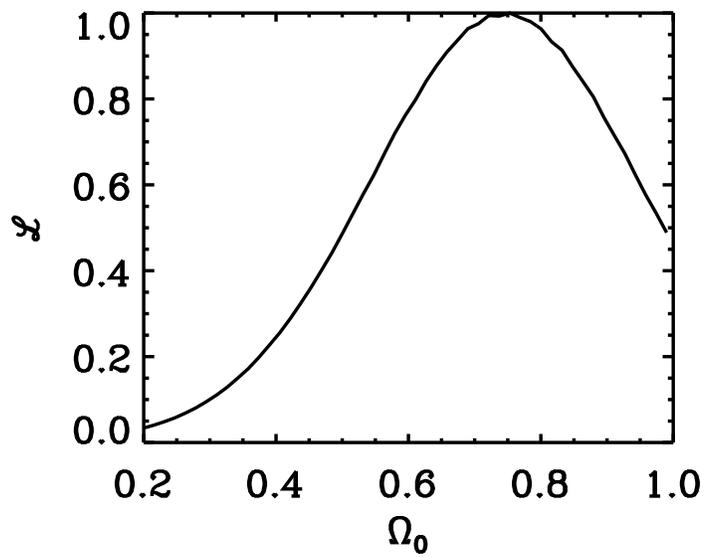}}}
~\vspace*{-1.cm}

\caption{ The likelihood function normalized to one
 obtained from the analysis of the relative abundance of clusters between $z = 0.05$ and $z=0.33$.
This function clearly favors a high density universe.
}
\end{figure}
 
\begin{center}
\textbf{References}
\end{center}

{\scriptsize
\noindent
\textbf {Adami C., Mazure A., Katgert P. , Biviano A.} 1998, A\&A, 336, 
63--82.\\
\textbf {Bartlett, J.G.} 1997, Proceedings of the 
1st Moroccan School of Astrophysics, 
ed. D.~Valls-Gabaud et al., A.S.P. Conf. Ser.,
vol. 126, p. 365--386\\
\textbf {Davis, M.} 1998, Proc. Natl. Acad. Sci., United Sates, 95, 78\\
\textbf {Dekel, A.} 1994, ARA\&A, 32, 371--418\\
\textbf {Ebeling, H.,   Edge, A.C., Fabian,  A.C.,  Allen, S.W.,  Crawford, C.S.} 1997, ApJL, 479, L101--104\\ 
\textbf {Eke, V.R., Cole, S., Frenk, C.S., Henry, J.P.} 1998,  to appear in MNRAS, 
astro-ph/9802350\\
\textbf {Evrard, A.E.} 1989, ApJL, 341, L71--74\\
\textbf {Henry, J.P.}  1997, ApJL, 489, L1--5\\
\textbf {Oukbir, J.,  Blanchard A.} 1992, A\&AL, 262, L21--24\\  
\textbf {Press W.H., Schechter, P.} 1974, ApJ, 187, 425--438 \\
\textbf {Riess, G.A., et al.} 1998,  Accepted to the Astronomical Journal, astro-ph/9805201\\
\textbf {Sadat, R., Blanchard, A., Oukbir, J.} 1998, A\&A,  329, 21--29\\
\textbf {Viana, P.,  Liddle, A.R.} 1998, submitted to MNRAS, astro-ph/9803244\\
}
\end{document}